# A Method for Determination of the Reynolds Stress and Turbulence Structure in Wall-Bounded Flows

T.-W. Lee

*Mechanical and Aerospace Engineering, SEMTE, Arizona State University, Tempe, AZ, 85287*

**Abstract-** Using the Lagrangian transport of momentum, the Reynolds stress can be expressed in terms of basic turbulence parameters. The Reynolds stress gradient represents the lateral transport of streamwise momentum, balanced by the $u'^2$ transport, pressure and shear force terms in the momentum equation. Data from direct numerical simulations (DNS) have been used to validate this approach, with a good degree of consistency and agreement. This expression for the Reynolds stress gradient adds a transport equation to the Reynolds-averaged Navier-Stokes (RANS) equation, so that the turbulence structure can be determined if the third unknown variable, $u'^2$, can be deteremined. Here, we consider alternative structure-based methods for reconstruction of $u'^2$ profiles for wall-bounded flows. The input of $u'^2$ profile, in conjunction with the two transport equations (RANS and the current Reynolds stress equation), allows for full reconstruction of the turbulent structure in wall-bounded flows, with reasonable agreement with DNS data.

T.-W. Lee
Mechanical and Aerospace Engineering, SEMTE
Arizona State University
Tempe, AZ 85287-6106
Email: attwl@asu.edu

**INTRODUCTION**

Determination of the Reynolds stress in terms of root turbulence parameters has profound implications in fluid physics and engineering, as it prescribes the mean flow structure through the Reynolds-averaged Navier-Stokes (RANS) equation. As is well known, turbulence models attempt to express the Reynolds stress in terms of known (or computable) parameters so that artificial closure is achieved. The procedure is typically to expand the Reynolds stress or the turbulence viscosity, in terms of higher-order terms, e.g. turbulence kinetic energy, dissipation and third-order correlations in the case of Reynolds stress models. In turn, these higher-order terms need to be algebraically modelled as reviewed in some articles [1-6].

Recently, we have developed a new formulation for the Reynolds stress based on the Galilean-transformed Navier-Stokes equation [7-9], where a simple, explicit expression is derived for the Reynolds stress in terms of root turbulence variables as shown below:

$$\frac{d(u'v')}{dy} = -C_1 U \left[\frac{d(u'^2)}{dy} + \frac{1}{\rho}\frac{d|P|}{dy}\right] + \frac{1}{\nu}\frac{d^2 u'}{dy^2} \tag{1}$$

The derivation can be found in Refs. 5-7. Eq. 1 is an expression arising from the momentum balance for a control volume moving at the local mean velocity, where the Reynolds stress represents the lateral (y-direction) transport of u' momentum [7-9]. This lateral transport is balanced by the longitudinal transport and force terms on the right-hand side (RHS) of Eq. 1, where the d/dx gradients are transformed to d/dy, following the displacement of the fluid [9]. Viscous shear stress due to gradient of the turbulent fluctuation velocity (u' = u'$_{rms}$) is a new concept;

however, we can visualize that if there exists a gradient in the mean fluctuation velocity then it will lead to shear stress in the mean in this moving coordinate frame. Once we have the gradient from Eq. 1, then the Reynolds stress itself can be obtained through numerical or algebraic integration. For example, for rectangular channel flows we can integrate Eq. 1 by parts:

$$<u'v'> = -C_1 \left[ Uu'^2 - \int_0^y u'^2 \frac{dU}{dy} dy \right] + \frac{1}{v} \frac{du'_{rms}}{dy} \qquad (2)$$

We can see that the Reynolds stress contains a non-local integral term, indicating a cumulative effect of the mean velocity gradient. This points to the misfit with linear, localized dependence on the mean velocity gradient in k-ε type of turbulence models. Similar observation for the necessity of non-local eddy viscosity has been made by Hamba [5]. Algebraic Reynolds stress models use power expansions in terms of the strain and vorticity [6]. Eqs. 1 and 2 indicate that the basis functions, strain and vorticity, used in this type of expansions may have difficulty in obtaining generalized expressions for the Reynolds stress since the physical terms (in Eqs. 1 and 2) leading to the Reynolds stress do not have any apparent dependence on these basis functions. Instead, if the agreement between Eq. 1 (or Eq. 2) and turbulence data could be trusted, as illustrated in our earlier work [1-5] and below, then it may be simpler and more direct to use Eq. 1 or 2 as the algebraic Reynolds stress "model" equation. Alternatively, for a generalized expression for the Reynolds stress, one may seek an expansion for the *gradient* of $R_{ij}$ (e.g. u'v') in terms of various gradient of the diagonal components, $R_{ii}$ (e.g. $u'^2$, $v'^2$), as similar to the representation in Eq. 1. In particular, since $R_{ii}$ components are the missing elements in our current approach, merging of the current method with algebraic Reynolds stress modeling may offer an interesting opportunity for further work, aspects of which are currently being investigated.

The effectiveness of the current approach in determining the Reynolds stress gradient has been validated for canonical flows [7-9]. Here, for further clarification we can examine the break-down of the terms contributing to the Reynolds stress gradient, i.e. $u^2$ transport, pressure and viscous force terms on the RHS of Eq. 1, as plotted in Fig. 1. This is a much simpler Reynolds stress (gradient) budget than the Eulerian counterparts with many more terms [3]. We can see that the viscous term is small, but affects the momentum transport near to the wall. The longitudinal transport term has the largest effect, particularly near the wall due to large $u'^2$ gradient. The pressure force is pervasive, and without accounting for it correct momentum balance will not be achieved. The pressure term is derived from $-\rho v^2$ for channel flows [10]. These three terms all add up to the Reynolds stress gradient, nearly precisely, as shown in Fig. 1, where Reynolds stress gradient computed using Eq. 1 is compared with direct numerical simulation (DNS) data of Hopkins et al. [11] for channel flow at $Re_\tau = 1000$. For large Reynolds numbers, all of the above effects have significant influence near the wall (y/d < 0.06). Beyond that, the Reynolds stress gradient has a small positive value, leading to gradual increase toward the centerline value for the Reynolds stress. For flow over flat plate at zero (ZPG), Eq. 1 again agrees well with DNS data of Spalart [9] for Re = 670 in Fig. 2. For the adverse pressure gradient, Eq. 1 us compared with experimental data by Kaehler [10], again favorably in Fig. 3. For APG, we integrate the Reynolds stress gradient, so that it can be directly compared with the Reynolds stress data [10]. These aspects of the current approach, from the dynamic derivation to the comparisons with various data sets, should point to the validity of this method.

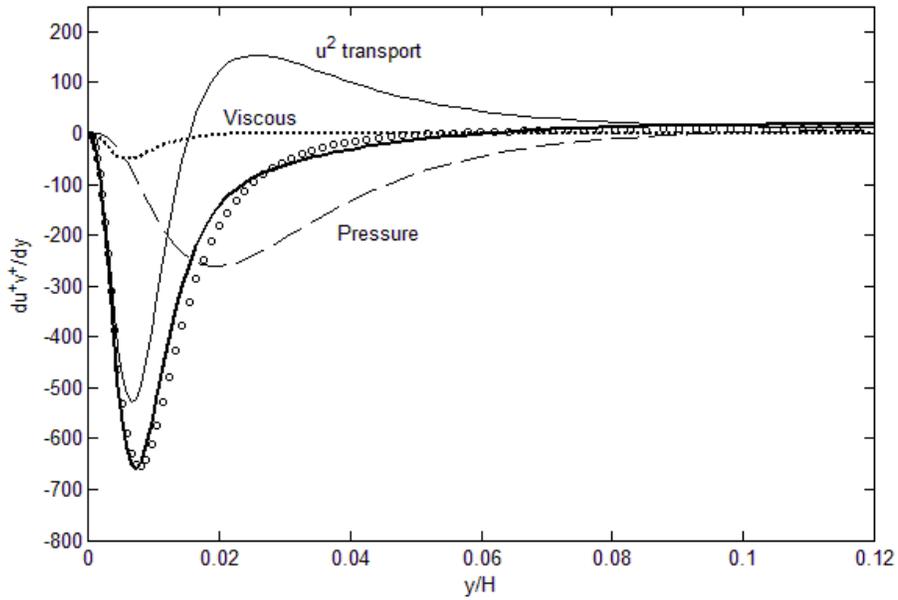

**Fig. 1.** Reynolds stress gradient budget for channel flows, consisting of u'$^2$ transport, viscous and pressure terms in Eq. 1. DNS data is from Graham et al. [11].

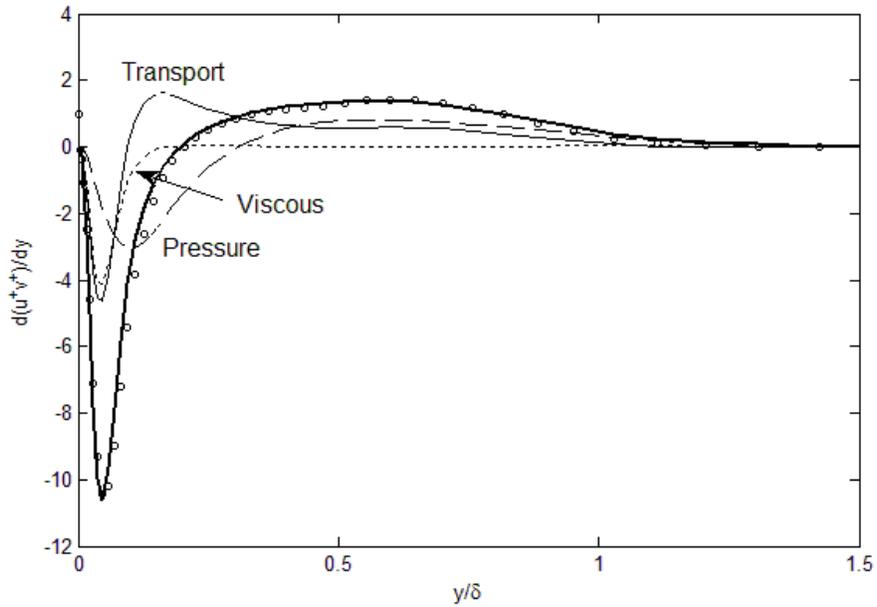

**Fig. 2.** Reynolds stress gradient budget for flow over a flat plate at zero pressure gradient. DNS data is from Spalart [12].

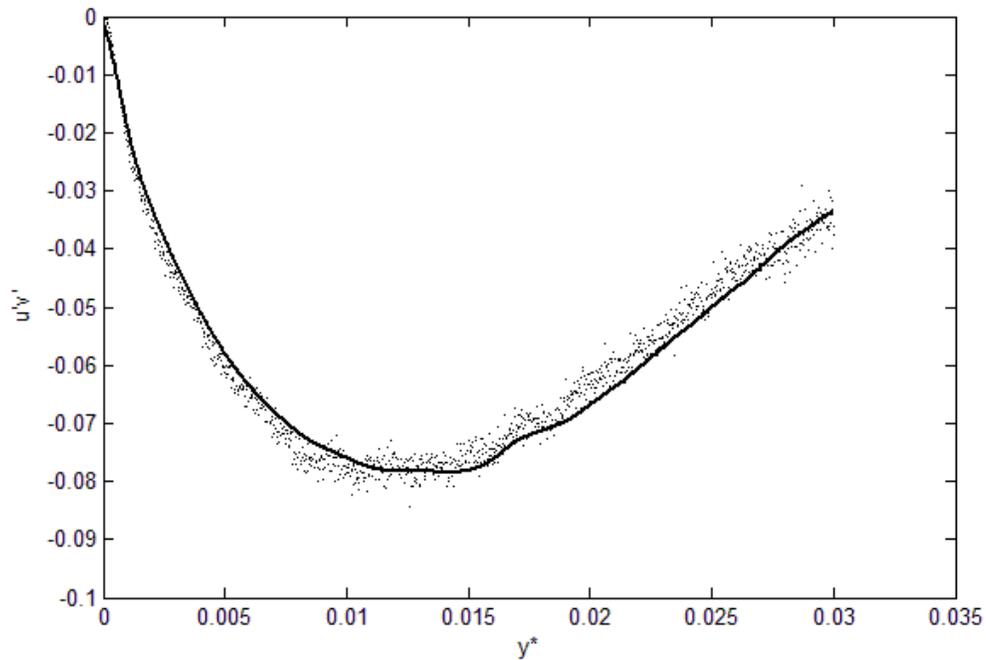

**Fig. 3. Reynolds stress in boundary layer flow over a flat plate with adverse pressure gradient. Line is from Eq. 1, and the data are from Bross et al. [13].**

The Reynolds stress gradient budget in Figs. 1-2 illustrates that the origin of the Reynolds stress (gradient) is the cross-stream transport of turbulent fluctuation momentum, to balance the longitudinal transport and the sum of pressure and viscous forces acting on the fluid. When viewed from an observer travelling at the local mean velocity, only transport of streamwise momentum is due to the $u'^2$, while $u'v'$ represents that cross-stream transport. The net transport of these momentum fluxes is balanced by the force terms, which are the pressure and viscous forces. As noted earlier determination of the Reynolds stress is the fundamental requirement for solving the turbulence problem. However, this approach still requires an input of key turbulence variable, $u'^2$, in Eq. 1. It is possible to write the transport equation for $u'^2$, in Lagrangian or Eulerian coordinate frame, similar to the transport equation for turbulent kinetic energy in k-ε models [1]. However, this will introduce higher-order terms, requiring modeling. If $u'^2$ is available through an independent equation or method, then the Reynolds-averaged Navier-Stokes equation (RANS) along with Eq. 1 can be directly integrated, or iteratively solved. For example, in wall-bounded flows, RANS may be simplified or approximated as:

$$\mu \frac{d^2 U}{dy^2} = \frac{dP}{dx} + \rho \frac{d(u'v')}{dy} \qquad (3)$$

Therefore, the aim of this work is to examine feasible alternate methods to compute or estimate u'$^2$ profile, so that turbulence structures including the mean velocity and the Reynolds stress can be reconstructed with minimal empiricism in wall-bounded flows. As noted earlier, we can use Eq. 1 or 2 within an algebraic Reynolds stress model framework, which is currently being considered in a separate study. Here, however, we investigate the reconstruction methods for u'$^2$, based on structural constraints. These methods will be discussed below, and the results compared with DNS results.

**RECONSTRUCTION METHODS**

The discussion above points to the need for the u'$^2$ variable in the current formulation (Eq. 1). One simple method is to use existing turbulence models for turbulent kinetic energy, k, and extract u'$^2$, which can then be input into Eq. 1 for the Reynolds stress. The Reynolds stress profile can then be cycled back to the turbulent kinetic energy equation in the turbulence model to update k and u'$^2$ until convergence. The Reynolds stress from Eq. 1 can then be used to directly compute the mean velocity profile through Eq. 3. This route also points to embedding Eq. 1 in existing turbulence models, as a physics-based "model" for the Reynolds stress, similar to the algebraic

Reynolds stress modeling [6]. Eq. 1 would then replace the algebraic model for the $R_{ij}$, while $R_{ii}$ terms are still computed using the standard methods. This approach is currently also being investigated. An alternate class of methods is to use observational constraints, and reconstruct $u'^2$ profiles for a given Reynolds number. For example, let us examine $u'^2$ profile in wall-bounded flows (e.g. Fig. X) from a "structural" perspective. In Fig. 4, $u'^2$ profiles are plotted vs. y/H in turbulent channel flows, for $Re_\tau$ ranging from 130-650 [14], 1000 and 5200 [11]. We can identify some physical trends in the structure. The peak location moves closer to the wall as the Reynolds number increases. This peak shift toward the wall involves increase in the slope of the $u'^2$ profile, thereby increasing the total dissipation, defined as:

$$\varepsilon = \int_0^1 \left(\frac{du'^+}{dy}\right)^2 d(\frac{y}{d}) \qquad (2)$$

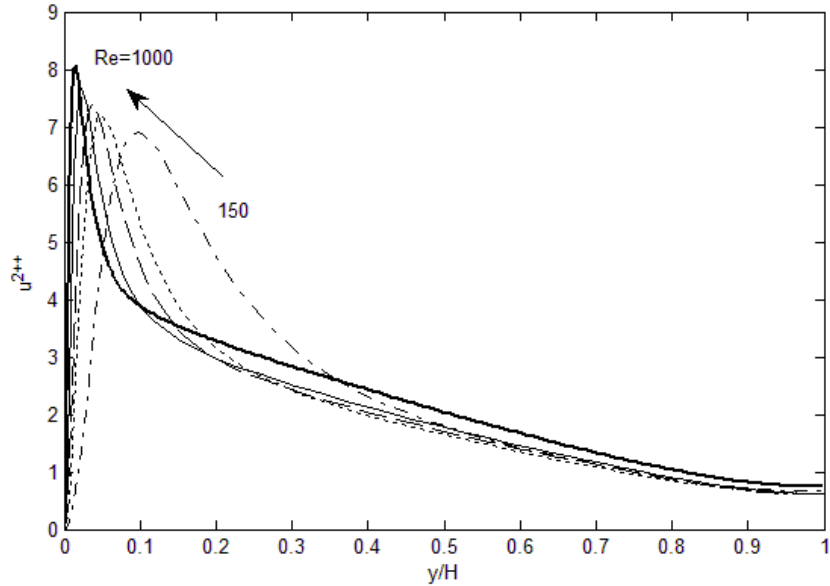

**Fig. 4. Reynolds stress gradient budget for channel flows, consisting of $u'^2$ transport, viscous and pressure terms in Eq. 1. DNS data is from Graham et al. [11].**

If we invoke the basic definition of entropy as the degree of disorder in the system, similar to the Shannon's entropy [15], then this behavior can be interpreted as the total dissipation increasing to achieve the maximum entropy allowed by the Reynolds number. The higher the Reynolds number, the restoring effect of viscosity is reduced proportionately so that higher entropy state can be attained. We can confirm this increase in dissipation, $\varepsilon$, in Fig. 5, which shows $\varepsilon$ increasing linearly as a function of $Re_\tau$, The $\varepsilon$ data in Fig. 5 were obtained from two different DNS data sets of Iwamoto [14] for $Re_\tau = 110 \sim 650$ and Graham et al. [11] for $Re_\tau = 1000, 5200$, respectively. Even across two different data sets, the linearity in the dissipation is retained. The peak location, $(y/d)_{peak}$ also scales in a predictable manner as a function of the Reynolds number (Fig. 6). In addition, $u'^2$ integrated over the half-width (= E) is very close to being constant, when normalized by the friction velocity ($u_\tau$), as shown in Fig. 5.

$$E = \int_0^1 u'^{+2}(y) d(\frac{y}{d}) \qquad (4)$$

Thus, for a constant E the longitudinal turbulent kinetic energy, $u'^2(y)$ must distribute itself in space to attain the maximum allowable dissipation, $\varepsilon$, at a given Reynolds number, leading to the increasingly skewed profiles as shown in Fig. 4. Similar observations for boundary layer flows over a flat plate can also be made, and summarized as follows:

(1) Integrated turbulent kinetic energy, $E_x$, stays constant when the turbulent kinetic energy is normalized by the friction velocity: $E_x$ = constant.

(2) The location of the peak turbulent kinetic energy is inversely proportional to the Reynolds number: $(y/d)_{peak} \sim Re_\tau^{1/2}$.

(3) Integrated dissipation, e, increases linearly with the Reynolds number. $\varepsilon \sim Re_\tau$.

(4) The ratio, $\varepsilon_1/\varepsilon_2$, increases with $Re_\tau$, where $\varepsilon_1 = \int_0^{(\frac{y}{d})_{peak}} \left(\frac{du'^+}{dy}\right)^2 d(\frac{y}{d})$ and $\varepsilon_2 = \int_{(\frac{y}{d})_{peak}}^1 \left(\frac{du'^+}{dy}\right)^2 d(\frac{y}{d})$.

(5) $v'^2$ has a log-normal distribution, with its dissipation, $\varepsilon_y$, increasing linearly with $Re_\tau$.

(6) Boundary conditions are U=u'v'=u'^2= 0 at y =0, and dU/dy=d(u'v')/dy=du'^2/dy = 0 at the centerline (channel) and y >> 1 (boundary layer flow).

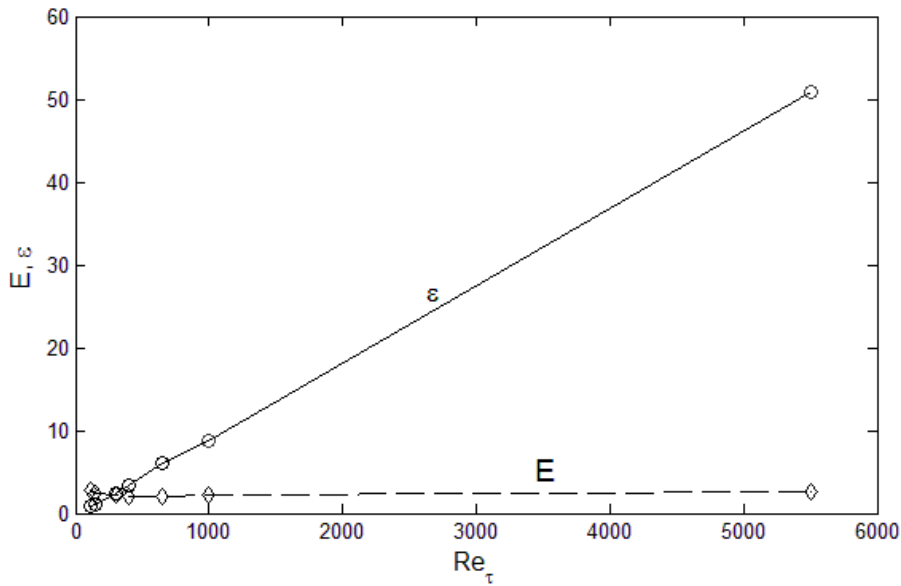

**Figure 5. Total integrated $u'^2(E)$ and dissipation ($\varepsilon$) as a function of the Reynolds numbers, from the DNS data [11, 14].**

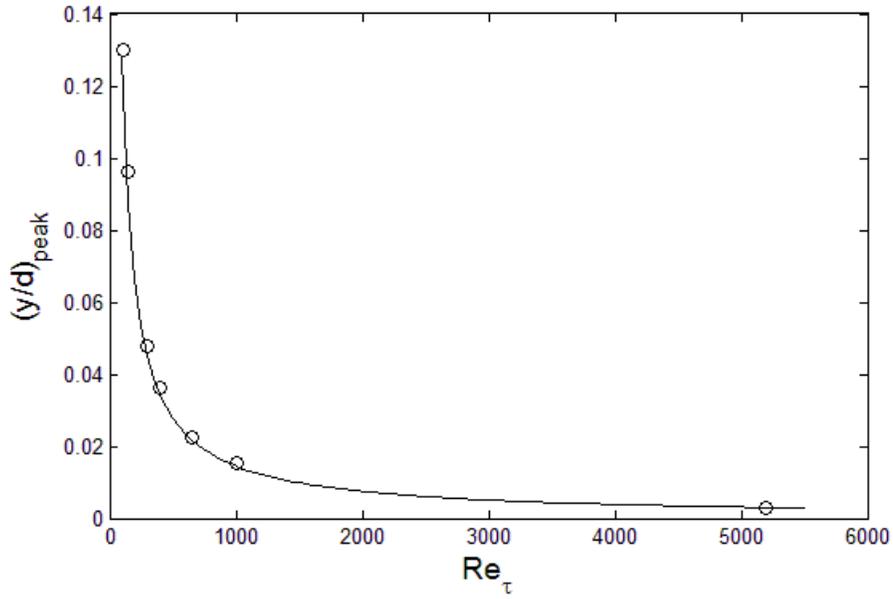

**Figure 6. The location of the $u'^2$ peak as a function of the Reynolds numbers, from the DNS data [11, 14].**

The observations above indicate that $u'^2$ profiles follow certain physical constraints as imposed by the conservation of energy, maximum achievable dissipation (entropy), and the boundary conditions. These are some necessary conditions for reconstructing $u'^2$ profiles. These conditions (or constraints) vastly reduce the degree of freedom in the $u'^2$ profiles, allowing for iterative reconstruction along with the other transport equations (Eqs. 1 and 3). These structural properties are used to infer the initial spatial distribution of $u'^2$, and iteratively determine its final structure. One template for the initial $u'^2$ profile is to use lognormal (for inner) and beta functions (for outer), constrained by the above conditions. An alternate method is to use a four-term Fourier series for $du'^2/dy$ for a given Reynolds number, and "stretch" this sampled Fourier series to achieve the total dissipation, $\varepsilon$, for different Reynolds numbers, while again obeying all of the other constraints described above. These function and Fourier-series methods for $u'^2$ reconstruction are described further in the Appendix. The iterative algorithm is shown in Fig. 7, which consists of initial estimate for $u'^2$ profiles, computing the $u'v$ and U profiles using Eqs. 1 and 3, then iterating the $u'^2$ function parameters until the profiles converge while satisfying all the boundary conditions and piecewise smoothness. Further details of the procedure, including lognormal behavior of $v'^2$ and $w'^2$ profiles, are provided in the Appendix. For moderate Reynolds number up to $Re_\tau = 1000$, the number of iterations was manageable, to the extent that sufficient accuracy in $u'^2$ gradient is achieved for the rest of calculation as demonstrated below. In summary, Eqs. 1, 3 and the constraints (as the third "equation") furnish us with three equations, to iteratively solve for three unknowns, U, u'v' and $u'^2$.

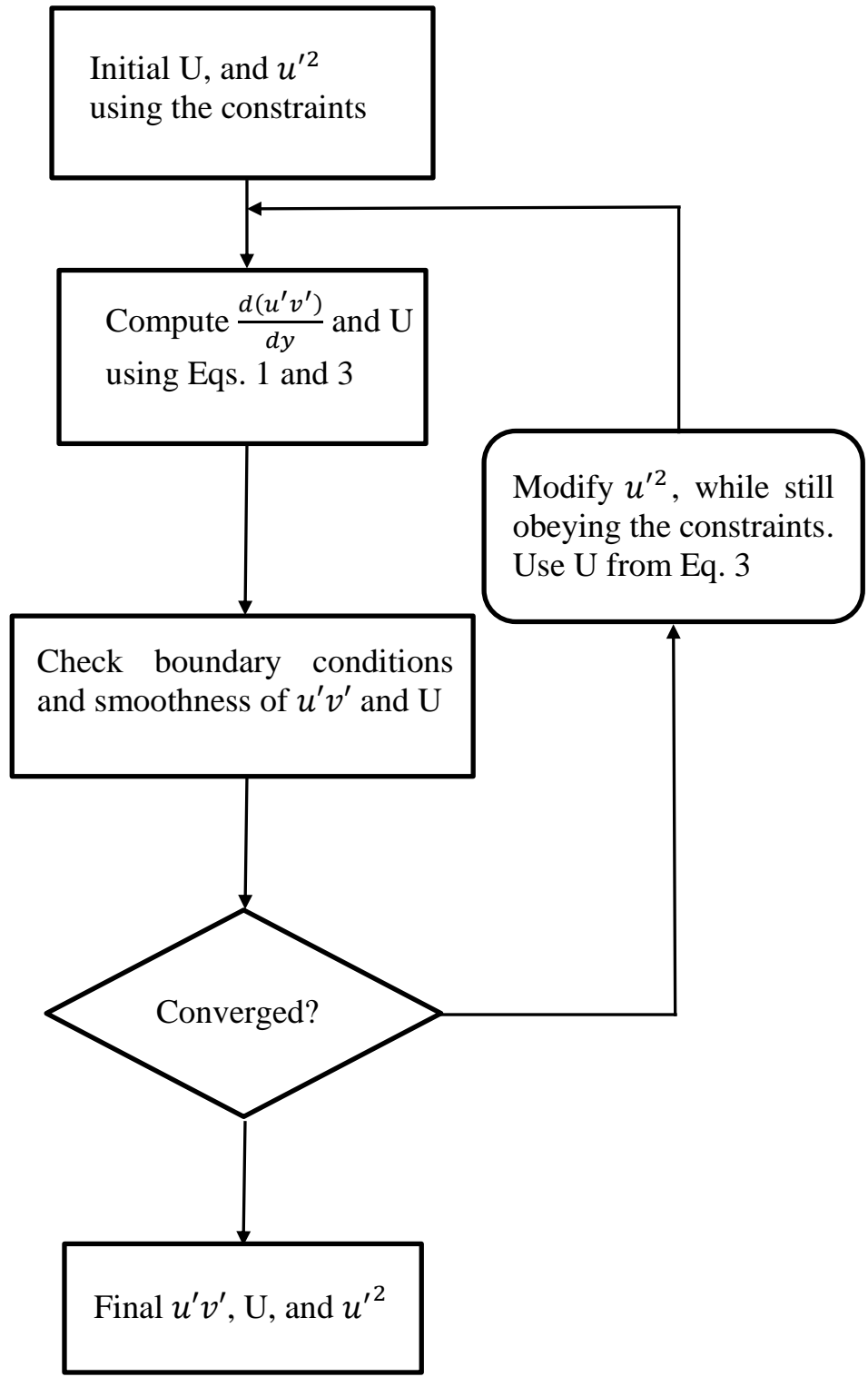

**Figure 7. A flow chart for the iterative solution method to determine the turbulence structure in wall-bounded flows.**

**RESULTS AND DISCUSSION**

The methods discussed in the previous section lead to $u'^2$ profiles, which are compared with DNS data in Fig. 8 for channel flows. We can see that fairly close $u'^2$ profiles are achieved in comparison to the DNS data, when the above iterative algorithm is used. The key constraint is the location of the maximum $u'^2$ (shown in Fig. 6), which also corresponds to the location for $du'^2/dy$. Once this is set, then the remaining profile only has limited degree of freedom, and the final form for the $u'^2$ is reached with a small number of iterations. For the boundary layer flow over a flat plate, we use the Fourier-series method for $du'^2/dy$, since the inner and outer shape of this profiles resembles a Fourier series. Also, we note that the Reynolds stress gradient is directly related to $du'^2/dy$, so that this parameter needs to be reconstructed with a fair amount of accuracy. Again, $u'^2$ profile that is fairly close to the DNS data is arrived by using the constraints, as shown in Figs. 8 and 9. Thus, following the structural properties and a small number of iterations, $u'^2$ and $du'^2/dy$ profiles can reasonably be approximately converged with the current iterative solution procedure.

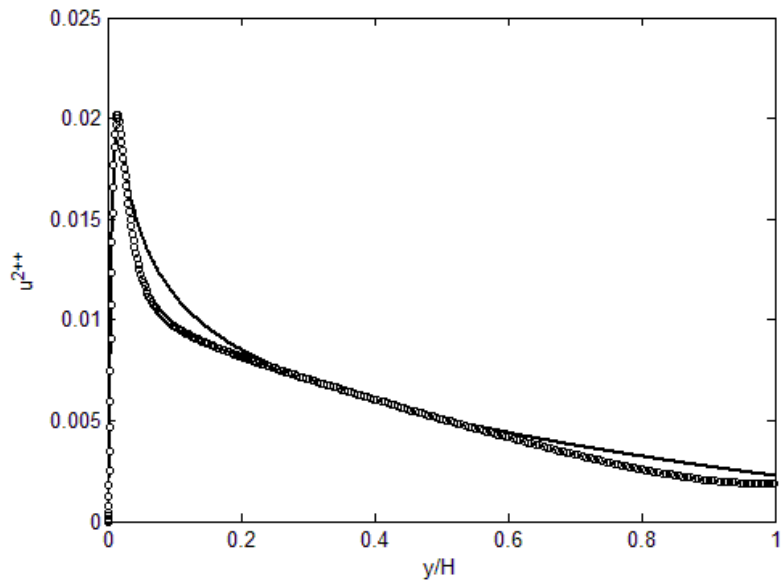

**Fig. 8(a). Reconstructed $u'^2$ profile, by integrating the gradient in Fig. 8a. Comparison is made with the DNS data (symbols) by Graham et al. [11].**

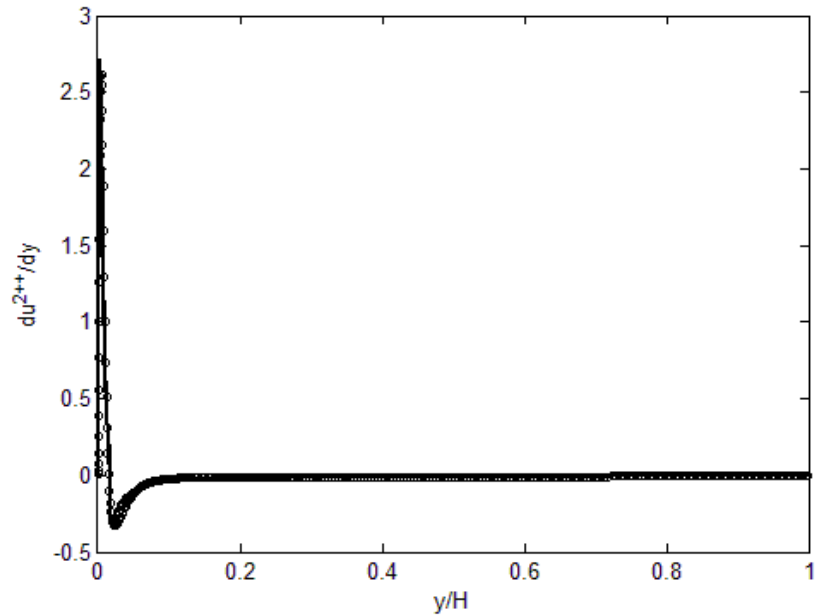

**Fig. 8(b).** Reconstructed u'² gradient for channel flows, using the structural constraints and iteration. Comparison is made with the DNS data (symbols) by Graham et al. [16], by finding the numerical derivative of u'² profile.

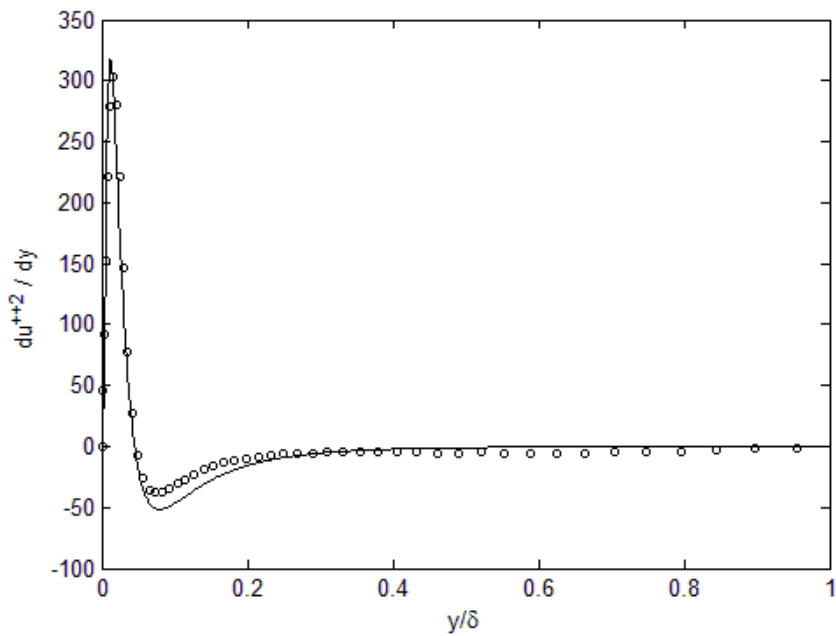

**Fig. 9(a).** Reconstructed u'² gradient for boundary layer flow over a flat plate, using the structural constraints and iteration. Comparison is made with the DNS data (symbols) by Spalart [12], by finding the numerical derivative of u'² profile.

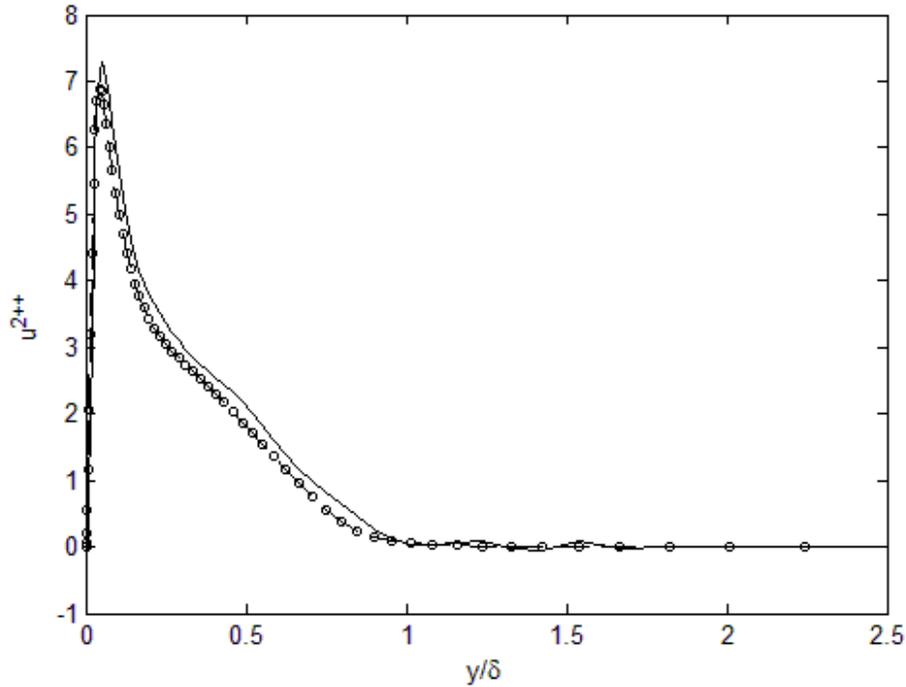

**Fig. 9(b). Reconstructed u'$^2$ profile, by integrating the gradient in Fig. 9a. Comparison is made with the DNS data (symbols) by Spalart [12].**

Eqs. 1 and 3, along with constraints for u'$^2$ thus constitutes a set of iteratively solvable constrained equations for u'$^2$, U and u'v', where u'$^2$ is the key variable in Eq. 1, leading to the Reynolds stress then to the mean velocity profile through RANS (Eq. 3). We can now examine the mean velocity and Reynolds stress profiles, for channel flows (Figs. 10 and 12) and boundary layer flow over a plate (Figs. 11 and 13), obtained in this manner. The results are again compared with DNS data [11] with decent agreements with the current method for both the channel and boundary layer flows. The discrepancy away from the wall is evidently from the accumulation of errors during numerical integration of Eq. 1, starting from the wall. In addition, any errors associated with reconstruction of u'$^2$ profiles will also be reflected in subsequent calculations in Eqs. 2 and 3. Nonetheless, current approach constitutes a physics-based algorithm based on Lagrangian momentum transport, along with constrained iteration for $u'^2$, leading to direct computation of the Reynolds stress without any modeling. The results for the Reynolds stress for both channel and flat plate flows are fairly close to the DNS data, with errors subject to improvements depending on the reconstruction and iterative algorithm.

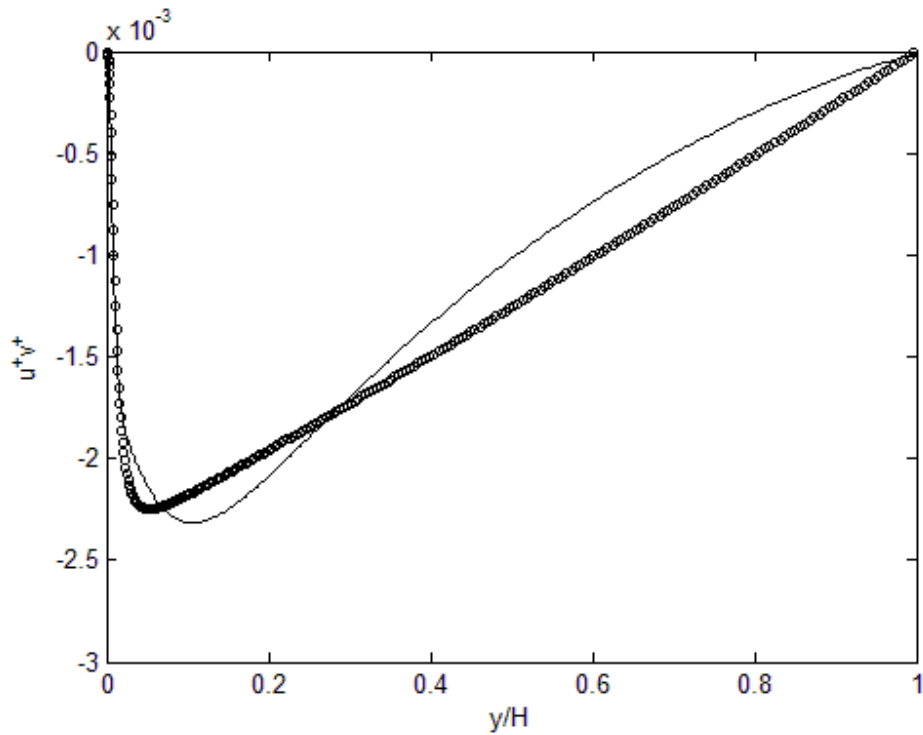

**Fig. 10. Computed Reynolds stress obtained by integrating Eq. 2, for channel flow. Comparison is made with the DNS data (symbols) by Graham et al. [11].**

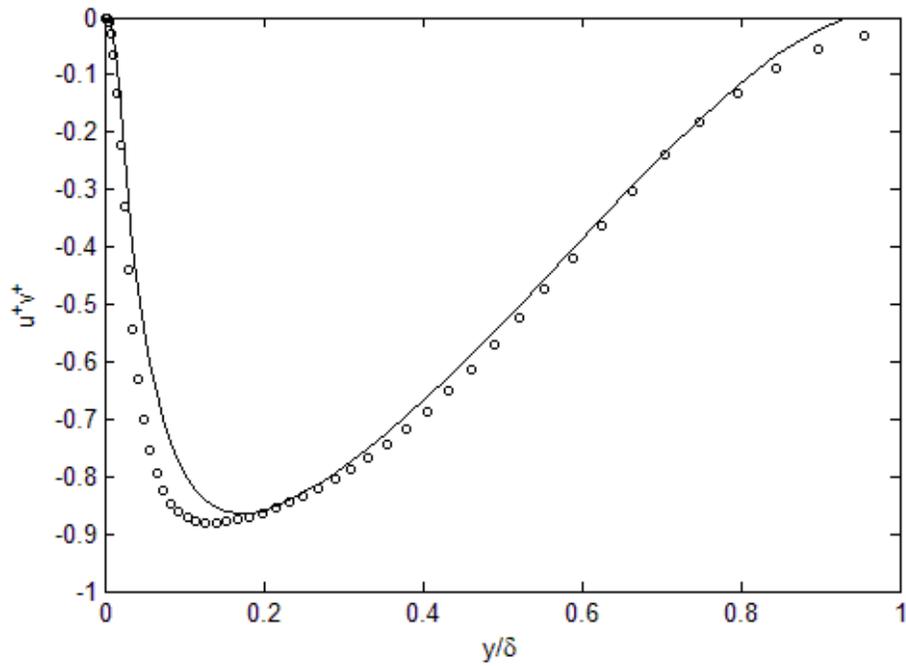

**Fig. 11. Computed Reynolds stress obtained by integrating Eq. 2, for boundary layer over a flat plate. Comparison is made with the DNS data (symbols) by Spalart [12].**

For the mean velocity profile, the Reynolds stress can be inserted into the RANS (Eq. 3), and integrated. In order to minimize accumulation of errors in the Reynolds stress, the integration is performed from the wall and the centerline (or freestream) separately. The results are shown Figs. 11(a) and (b) for the channel and flat plate boundary layer flow, respectively.

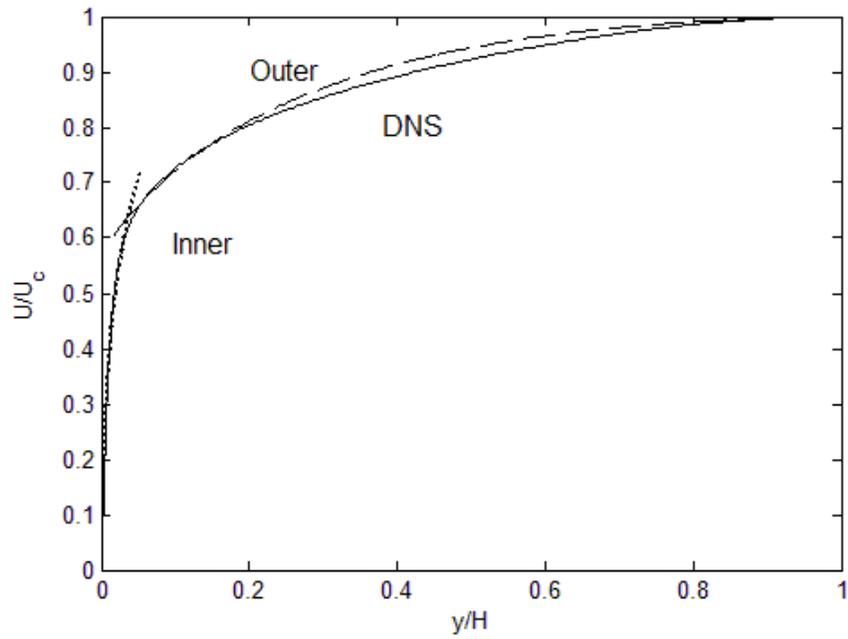

**Fig. 12. Computed mean velocity obtained by integrating Eq. 5, for channel flow. Comparison is made with the DNS data (symbols) by Graham et al. [11].**

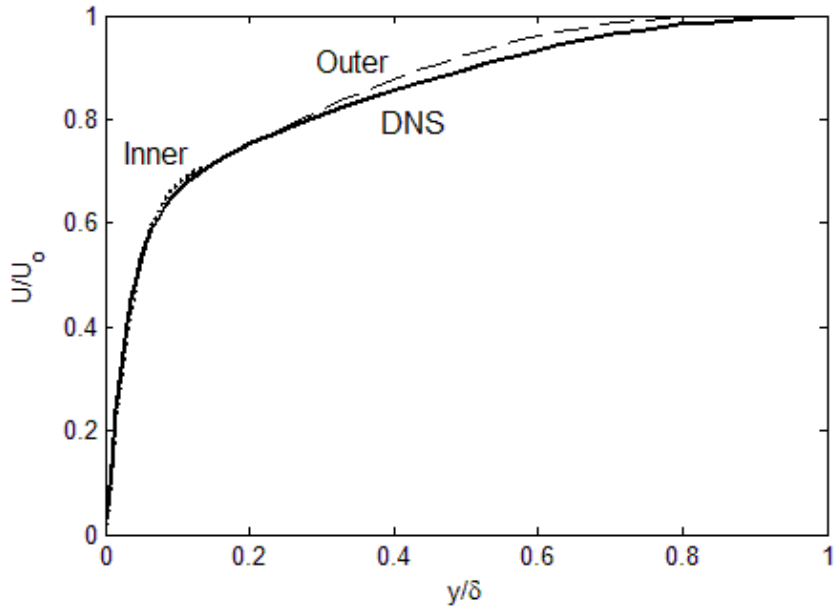

**Fig. 13. Computed mean velocity profile obtained by integrating Eq. 5, for boundary layer over a flat plate. Comparison is made with the DNS data (symbols) by Spalart [11].**

**CONCLUSIONS**

The Lagrangian transport analysis leads to an explicit expression for the Reynolds stress gradient in wall-bounded flows, which agree quite well with DNS data for channel and boundary layer flows. This transport equation (Eq. 1) makes it possible to determine the full turbulence structure, with some input for $u'^2$. We have used a maximum-entropy perspective in constrained iteration for constructing $u'^2$ profiles using its known structural properties and transport equations for the Reynolds stress (Eq. 1) and the mean momentum (RANS). The iterative algorithm is capable of generating plausible $u'^2$ profiles, which leads to the reconstruction of the mean velocity and Reynolds stress as well. As just noted, with some feasible input for $u'^2$ in Eq. 2, the transport equation for the Reynolds stress gradient (Eq. 1) furnishes an important step toward full reconstruction of the turbulent flow structure in wall-bounded flows.

**APPENDIX**

As noted eariler, we attempt two methods for reconstruction for $u'^2$ profile. The first is to use lognormal and beta functions for the inner and outer form for $u'^2$ profile; while the second is to use a "self-similarity" in the constraints, to stretch a sampled profile over a range of Reynolds numbers. The sampled profile is a four-term Fourier series, which satisfy all the constraints (1)-(5). The first method is relatively straight-forward in that lognormal and beta functions are applied for the inner and outer side of the $u'^2$ profile, respectively, where the inner and outer demarcation is the zero-crossing in the $du'^2/dy$, or the $y_{peak}/H$ location. The amplitude of the functions are iterated until the total energy, E, is achieved, while the function parameters are iterated to satisfy the dissipation, $\varepsilon$.

The second Fourier-series method involves the concept of self-similarity, not specifically in the spatial structure, but in the constraints. For example, linear increase in ε may be considered as a scaling behavior, due to its proportionality to the Reynolds number. Thus, a $u'^2$ profile is sampled at a given Reynolds number and written as a four-term Fourier series, again separately into inner and outer series. The demarcation point is again $du'^2/dy = 0$, or the $y_{peak}/H$ location  Then, the inner and outer series for $du'^2/dy$ can be stretched to achieve ε, at other Reynolds number, while adhering to all the other constraints.

As for the $v'^2$ and $w'^2$ profiles, they appear to follow the log-normal function fairly well, so that a single log-normal function following the respective constraints are used for $v'^2$ profile.

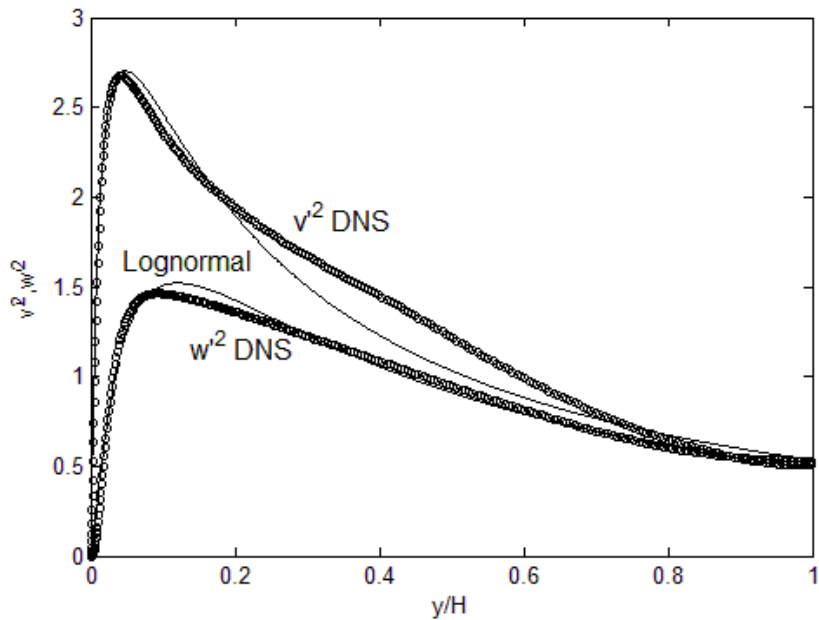

**Figure 14. A flow chart for the iterative solution method to determine the turbulence structure in wall-bounded flows.**

The dissipation, ε, and the zero-crossing point for the $u'^2$, $v'^2$, and $w'^2$, are shown in Figs. 15 and 16 below. We can see that that $v'^2$ profiles follow similar constraints as $u'^2$. This $v'^2$ profile is then used to obtain the pressure gradient term in Eq. 1, through $P=-\rho v'^2$, which follows from the y-component of the momentum equation for boundary layer flows [10].

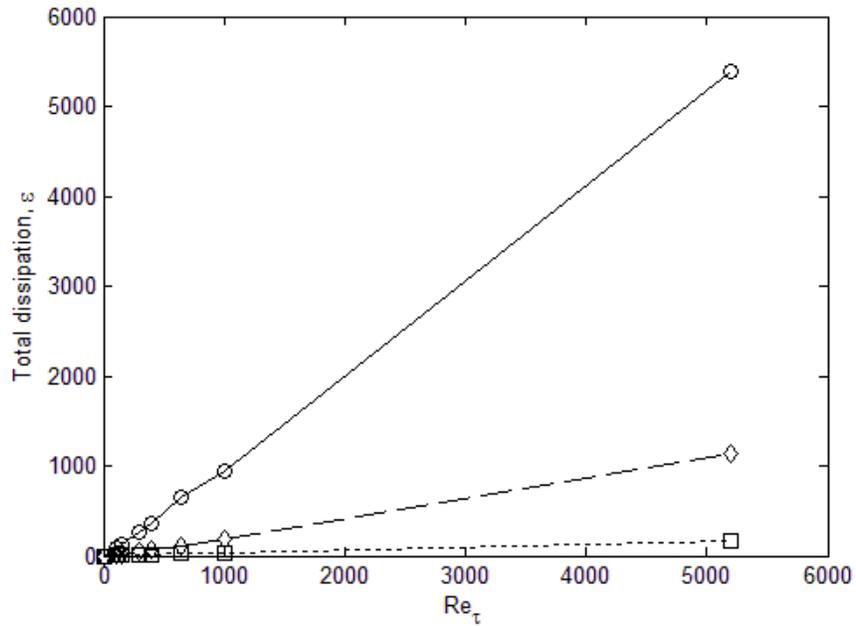

**Figure 15. A flow chart for the iterative solution method to determine the turbulence structure in wall-bounded flows.**

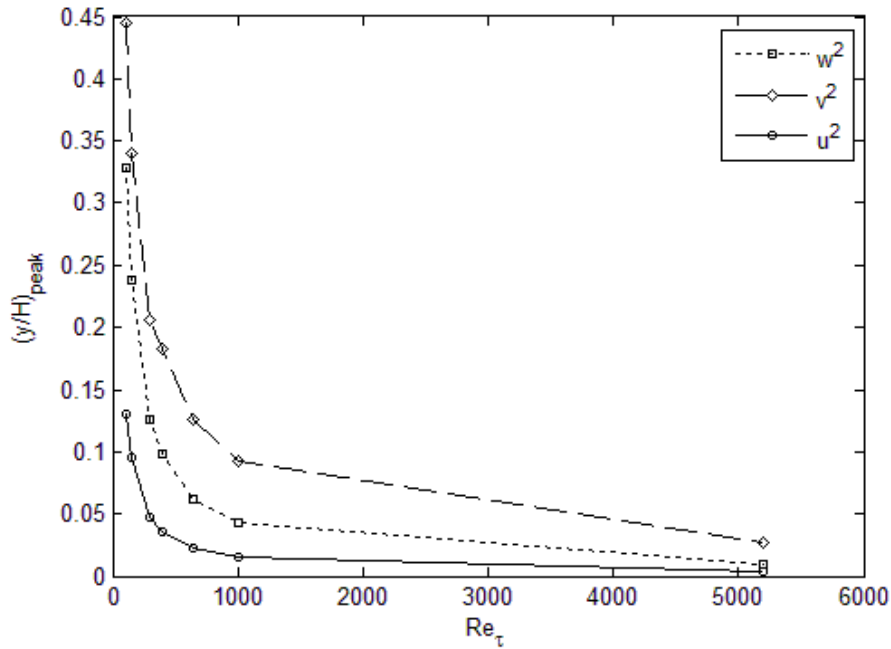

**Figure 16. A flow chart for the iterative solution method to determine the turbulence structure in wall-bounded flows.**